\documentclass{article}
\usepackage{spconf,amsmath,graphicx}
\usepackage{amsfonts}
\title{LIMITED-VIEW PHOTOACOUSTIC IMAGING RECONSTRUCTION WITH
DUAL \\DOMAIN INPUTS UNDER MUTUAL INFORAMTION CONSTRAINT
}
\name{Jiadong Zhang, Hengrong Lan, Changchun Yang, Shanshan Guo, Feng Gao, and Fei Gao$^*$}
\address{ Hybrid Imaging System Laboratory, School of Information Science and Technology,\\ShanghaiTech University, Shanghai 201210,China}
\begin{document}
%\ninept
%
\maketitle
\begin{abstract}
Based on photoacoustic effect, photoacoustic tomography is developing very fast in recent years, and becoming an important imaging tool for both preclinical and clinical studies.
With enough ultrasound transducers placed around the biological tissue, PAT can provide both deep penetration and high image contrast by hybrid usage of light and sound.
However, considering space and measurement environmental limitations, transducers are always placed in a limited-angle way, which means that the other side without transducer coverage suffers severe information loss.
With conventional image reconstruction algorithms, the limited-view tissue induces artifacts and information loss, which may cause doctors' misdiagnosis or missed diagnosis.
In order to solve limited-view PA imaging reconstruction problem, we propose to use both time domain and frequency domain reconstruction algorithms to get delay-and-sum (DAS) image inputs and k-space image inputs.
These dual domain images share nearly same texture information but different artifact information, which can teach network how to distinguish these two kinds of information at input level.
In this paper, we propose Dual Domain Unet (DuDoUnet) with specially designed Information Sharing Block (ISB), which can further share two domains' information and distinguish artifacts.
Besides, we use mutual information (MI) with an auxiliary network, whose inputs and outputs are both ground truth, to compensate prior knowledge of limited-view PA inputs.
The proposed method is verified with a public clinical database, and shows superior results with SSIM = 93.5622\% and PSNR = 20.8859.

\end{abstract}
\section{Introduction}
\label{sec:intro}
As an emerging non-invasive medical imaging technology, photoacoustic tomography (PAT) gains much attention in recent years.
Based on photoacoustic (PA) effect, PAT combines high contrast of optical imaging with deep penetration of ultrasound imaging \cite{C2}.
Excited by a laser pulse, the biological tissue emits ultrasound signals, which are detected by ultrasound transducer array \cite{wang2012photoacoustic}.
With these PA signals, some conventional reconstruction algorithms such as back projection and time reversal have been widely used to reconstruct PA imaging,
which shows distorted images with artifacts, especially in a limited-angle way.
The weakened texture information is mixed with artifacts and even covered by artifacts.
So how to distinguish texture information with artifacts and recover the weaken or unseen texture becomes the key challenge of limited-view PA imaging reconstruction.

With the popularity of deep learning, more and more researchers shift their work to learning-based PA reconstruction algorithms \cite{yang2020deep}\cite{hauptmann2020deep}.
Learning from supervised ground truth, deep neural networks can get prior knowledge to compensate the limited-view induced information loss.
However, these networks mostly take time domain images (e.g. DAS images) as inputs, which means networks can only distinguish texture information and artifacts from supervised labels.
It may cause reconstruction inefficiency or inadequacy.

To solve this problem, we propose to use both time domain and frequency domain reconstruction algorithms to get DAS images and k-space images as inputs.
The dual domain images share nearly same texture information but different artifact information \cite{spadin2020quantitative}.
Specifically, the proposed DuDoUnet with ISB takes dual domain inputs to further share information and distinguish texture and artifacts at input level.
There is another advantage to use k-space domain images: the artifact information is more concentrated in k-space,
which makes network easier to remove artifacts and reconstruct texture information \cite{lin2019dudonet}\cite{zhou2020dudornet}.

Besides, considering that neural network can only learn prior knowledge from supervised labels, we propose to use mutual
information as an auxiliary constraint to make sure that network's encoder can also learn enough prior knowledge.
We pre-train an auxiliary encoder-decoder architecture network, which takes ground truth as both inputs and outputs.
According to the basic idea of information bottlenecks \cite{slonim2000agglomerative}, encoder has compressed most of the useful information as latent representations to recover inputs.
These representations can make encoder of DuDoUnet learn more prior knowledge with mutual information constraint \cite{ahn2019variational}.

\section{Proposed method}
\label{sec:format}

\begin{figure*}[!t]
\centering
\includegraphics[width=7.1in]{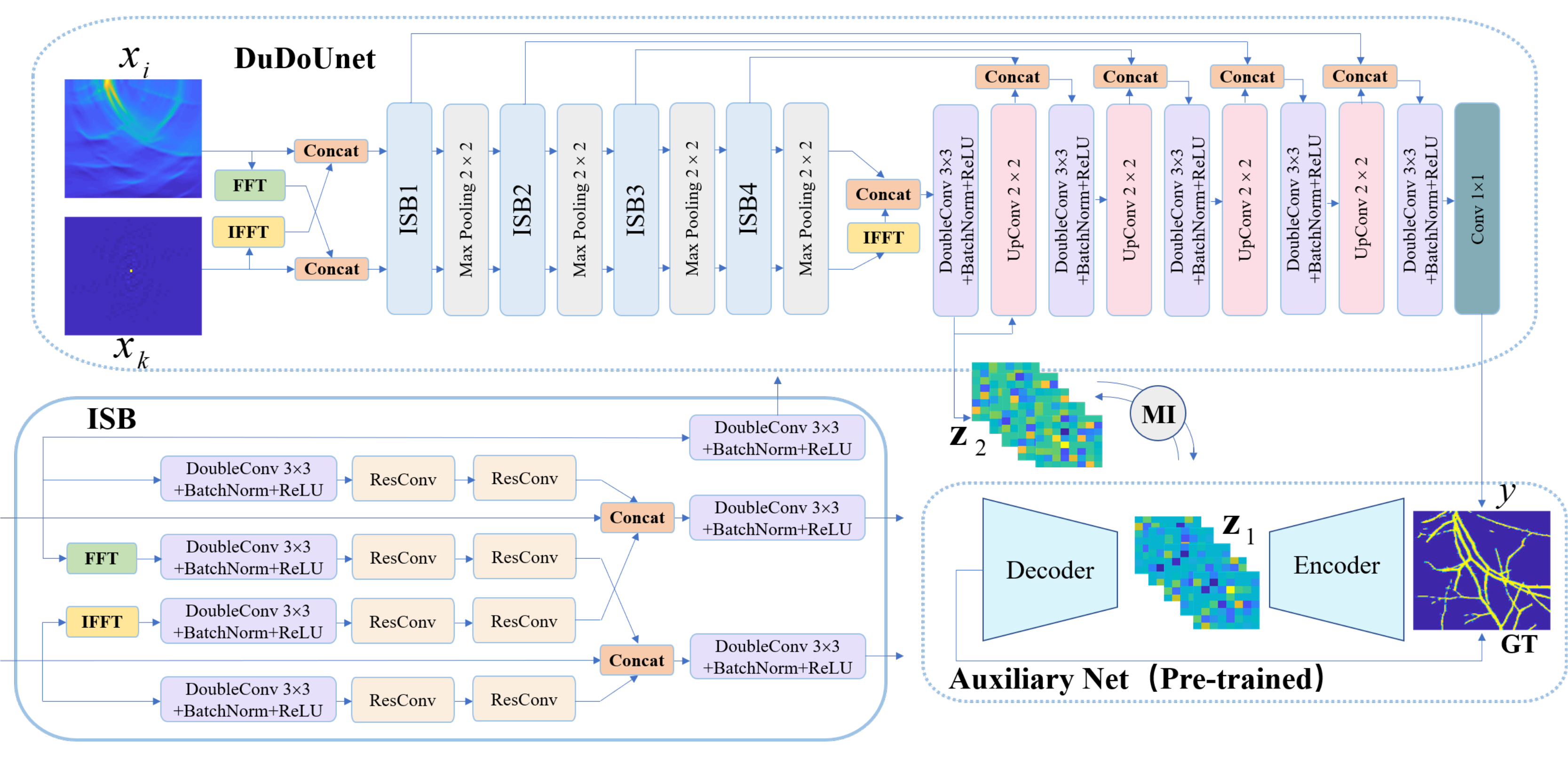}
\caption{The architecture of proposed network. DuDoUnet: Dual Domain Unet; ISB: Information Sharing Block; MI: Mutual Information; GT: Ground Truth.}
\label{fig:DuDoUnet}
\end{figure*}

In this section, we will make detailed description about our proposed network and method.
The overall network architecture is illustrated in Fig. 1, including the main DuDoUnet with ISB and a pre-trained auxiliary network.

\subsection{Dual Domain Unet}
The proposed DuDoUnet keeps most part of Unet \cite{ronneberger2015u}, but replaces convolutional layers of encoder with ISB.
ISB is specially designed for sharing dual domain inputs' vessel texture information and distinguishing them with artifacts.

ISB takes DAS image inputs to achieve two tasks.
One aims to get low-level features with convolutional layers.
The outputs skip to corresponding decoder layers.
The other task is to extract multi-level and dual domain features to achieve better performance:
(1) skipping without any computation to capture raw features;
(2) computed with deeper convolutional layers to extract high-level features;
(3) computed with fast fourier transform (FFT) and deeper convolutional layers to extract high-level frequency domain features.
The k-space image inputs perform just like the second task of DAS image inputs, but computed with inverse fast fourier transform (IFFT) in the third stream.
Both domains' three outputs are concatenated and computed with convolutional layers.
The final dual domains' outputs are the new inputs for next ISB after pooling layers.

The success of ISB counts as follow:
(1) multi-level features provide detailed textural information while keeping most input-level information;
(2) FFT and IFFT allow DAS and k-space image inputs sharing information without greatly influencing network parameters caused by different domain attributes.

The reconstruction log-likelihood loss function can be described as follow:
\begin{equation}
{\mathcal{L}_{{\text{recon}}}}(\theta ;{x_i},{x_k},y) = {\mathbb{E}_\theta }\,[log\ {\kern 1pt} {\kern 1pt} {f_\theta }(y|{x_i},{x_k})]
\end{equation}
where ${x_i}$, ${x_k}$ denotes DAS and k-space image inputs, $y$ denotes ground truth, $f(\cdot)$ denotes proposed DuDoUnet and $\theta $ represents all trainable parameters of DuDoUnet.
By maximizing ${{\cal L}_{{\rm{recon}}}}$, we can get a satisfactory  $\theta $ to perform reconstruction task.
In our experiment, we choose mean square error (MSE) as reconstruction loss function.

\begin{figure*}[!t]
\centering
\includegraphics[width=7.1in]{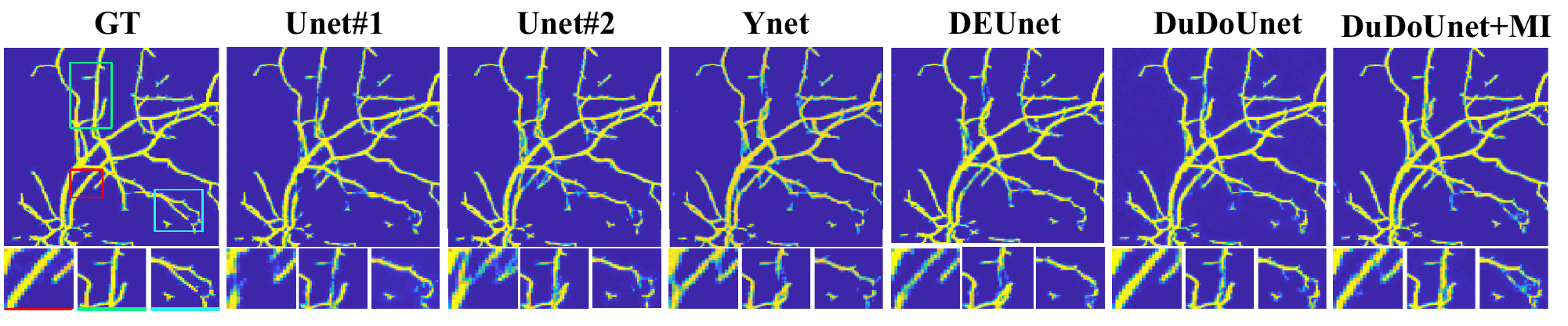}
\caption{Qualitative reconstruction performance of proposed method and other comparative methods. Three obviously different zones are enlarged below images.}
\label{fig:DuDoUnet}
\end{figure*}

\subsection{Prior Knowledge Compensation Based on Mutual Information}
As mentioned in introduction, we use an auxiliary network to provide latent prior knowledge representations ${{\bf{z}}_1}$.
Then mutual information is served as another constraint to guarantee that encoder of DuDoUnet can learn more texture information from ${{\bf{z}}_1}$,
rather only learn from supervised ground truth through many deep layers.
In order to show effectiveness of proposed method, we choose a normal autoencoder as auxiliary network in our experiment.

The new loss function is:
\begin{equation}
{\cal L}{\rm{ = }}{{\cal L}_{{\rm{recon}}}}{\rm{ + }}\lambda I({{\bf{z}}_1},{{\bf{z}}_2})
\end{equation}
where ${{\bf{z}}_2}$ denotes latent representations of DuDoUnet (equal to outputs of model's encoder), $I\left( {{{\bf{z}}_1},{{\bf{z}}_2}} \right)$
denotes mutual information between the pair of two latent representations. $\lambda $ is a positive penalty coefficient for regulation of mutual information term.

However, directly maximizing mutual information is intractable, so we use variational lower bound to replace mutual information term \cite{ahn2019variational}.
We use $p({{\bf{z}}_1}|{{\bf{z}}_2})$ to denote conditional distribution.
Because the true $p({{\bf{z}}_1}|{{\bf{z}}_2})$ is unknown, we define a variational distribution  $q({{\bf{z}}_1}|{{\bf{z}}_2})$ to approximate $p({{\bf{z}}_1}|{{\bf{z}}_2})$.
Then we can rewrite mutual information term as follow:
\begin{equation}
\begin{gathered}
\begin{gathered}
  I({{\mathbf{z}}_1},{{\mathbf{z}}_2}){\text{ = }}H({{\mathbf{z}}_1}) - H({{\mathbf{z}}_1}|{{\mathbf{z}}_2}) \hfill \\
  \quad \quad \quad \; = H({{\mathbf{z}}_1}) + {\mathbb{E}_{{{\mathbf{z}}_1},{{\mathbf{z}}_2}}}[log\;p({{\mathbf{z}}_1}|{{\mathbf{z}}_2})] \hfill \\
  \quad \quad \quad \; = H({{\mathbf{z}}_1}) + {\mathbb{E}_{{{\mathbf{z}}_1},{{\mathbf{z}}_2}}}[log\;q({{\mathbf{z}}_1}|{{\mathbf{z}}_2})] \hfill \\
  \quad \quad \quad \quad  + {\mathbb{E}_{{{\mathbf{z}}_1},{{\mathbf{z}}_2}}}[{\mathbb{D}_{KL}}(p({{\mathbf{z}}_1}|{{\mathbf{z}}_2})||q({{\mathbf{z}}_1}|{{\mathbf{z}}_2}))] \hfill \\
\end{gathered}
\end{gathered}
\end{equation}
where $H( \cdot )$ is information entropy and ${\mathbb{\mathbb{D}}_{KL}}(\cdot)$ denotes Kullback-Leiber divergence, which is non-negative.
Because auxiliary network is pre-trained, ${{\mathbf{z}}_1}$ is untrainable.
So we can directly maximize its variational lower bound log-likelihood  loss function $\tilde {\cal L}$:
\begin{equation}
\tilde {\cal L}{\text{ = }}{\mathbb{E}_\theta }[log{\kern 1pt} {\kern 1pt} {f_\theta }(y|{x_i},{x_k})]{\text{ + }}\lambda {\mathbb{E}_{{{\mathbf{z}}_1},{{\mathbf{z}}_2}}}[log\;q({{\mathbf{z}}_1}|{{\mathbf{z}}_2})]
\end{equation}

We employ Gaussian distribution to $q({{\mathbf{z}}_1}|{{\mathbf{z}}_2})$, which can be described as:
\begin{equation}
q({{\mathbf{z}}_1}|{{\mathbf{z}}_2}){\text{ = }}\frac{{\text{1}}}{{\sqrt {{\text{2}}\pi } \sigma ({{\mathbf{z}}_2})}}{e^{ - \frac{{{{({{\mathbf{z}}_1} - \mu ({{\mathbf{z}}_2}))}^2}}}{{2{\sigma ^2}({{\mathbf{z}}_2})}}}}
\end{equation}
where $\mu ({{\mathbf{z}}_2})$ and $\sigma ({{\mathbf{z}}_2})$ denote mean and variance of Gaussian distribution.
Assuming ${{\mathbf{z}}_1},{{\mathbf{z}}_2} \in {\mathbb{R}^{C \times H \times W}}$, then we can get $\mu ({{\mathbf{z}}_2}) \in {\mathbb{R}^{C \times H \times W}}$ with some convolutional computation.
But using same way to get $\sigma ({{\mathbf{z}}_2})$ will cause the instability of network, so instead we use following function:
\begin{equation}
\sigma ({{\mathbf{z}}_2}) = \frac{1}{{1 + {e^{\sum\limits_1^H {\sum\limits_1^W {{{\mathbf{z}}_2}} } }}}} + \varepsilon  \in {\mathbb{R}^C}
\end{equation}
where $\varepsilon $ is a positive constant to make sure network training is stable.
Now the variational mutual information constraint can be expressed as:
\begin{equation}
log\;q({{\bf{z}}_1}|{{\bf{z}}_2}){\rm{ = }} - \sum\limits_1^C {(\frac{{\sum\limits_1^H {\sum\limits_1^W {{{({{\bf{z}}_1} - \mu ({{\bf{z}}_2}))}^2}} } }}{{2{\sigma ^2}({{\bf{z}}_2})}}}  + log\;{\sigma}({{\bf{z}}_2}) + c)
\end{equation}
where $c$ is a constant and equal to $ \frac{{\log {\text{2}}\pi }}{2}$.

Equation (7) allows network constrains different channels with different weights.
It is a clear form for us to maximize the log-likelihood loss function.
It is worth noting that equation (7) becomes MSE when $\sigma ({{\mathbf{z}}_2})$ is equal to one.

\section{EXPERIMENT}
\label{sec:pagestyle}

We demonstrate proposed method with a public clinical vessel database \cite{staal2004ridge}.
The k-wave toolbox in MATLAB \cite{treeby2010k} is used to generate raw data from initial pressure.
128-channel linear-array transducers are placed above vessel images, which suffers typical limited-view PA imaging reconstruction problem.
Then we use both time domain and frequency domain reconstruction algorithms to get DAS and k-space image inputs from raw data.
We select 1500 samples to train our model and 500 samples to test.

In our experiment, we set $\lambda $  and $\varepsilon $  equal to one.
We evaluate the result with structural similarity index (SSIM) and peak signal-to-noise ratio (PSNR) quantitatively.
Considering our model takes dual domain inputs,
we choose Unet \cite{ronneberger2015u} and another two encoder architecture networks (Ynet \cite{lan2020net} and DEUnet \cite{wang2019dual}) as our comparative experiments
in order to prove the effectiveness of DuDoUnet with ISB.
Unet${\rm{\#}}$1 takes only DAS image inputs, and Unet${\rm{\# }}$2 takes dual domain inputs but with only one encoder.
Ynet and DEUnet take dual domain inputs with two separate encoders.
Because Unet${\rm{\# }}$2, Ynet and DEUnet do not have FFT or IFFT module, we additionally add a IFFT module
at k-space image inputs to guarantee comparison fairness of experiment performance.
The results are illustrated in Table 1.

From Table 1, we can see that Unet${\rm{\#}}$2 performs better than Unet${\rm{\#}}$1 with higher SSIM and PSNR.
It means that dual domain inputs indeed can improve reconstruction results.
By contrast, though Ynet and DEUnet takes dual domain inputs, the results cannot even surpass Unet${\rm{\#}}$1 due to their own network architecture.
In our method, DuDoUnet performs better than other comparative experiments, which means ISB can extract detailed and rich information with same inputs.
When adding variational mutual information constraint in training stage, we can get even better reconstruction results.

We illustrate the imaging results in Fig 2.
Three zones, which can obviously show reconstruction performance, are marked with boxes and enlarged below.
It is clear that all the comparative methods cannot reconstruct detailed information with destroyed vessel topological continuity or unwanted vessel structure.
By contrast, our models, especially DuDoUnet+MI, can recover vessel structure very well.

\begin{table}[!t]
\renewcommand{\arraystretch}{1}
% if using array.sty, it might be a good idea to tweak the value of
% \extrarowheight as needed to properly center the text within the cells
\caption{Comparison results of experiments}
\label{tab:result}
\centering
% Some packages, such as MDW tools, offer better commands for making tables
% than the plain LaTeX2e tabular which is used here.
\setlength{\tabcolsep}{5mm}{
\begin{tabular}{ccc}
\hline
Methods &SSIM($ \times {\rm{1}}{{\rm{0}}^{{\rm{ - 2}}}}$) & PSNR   \\
\hline
Unet${\rm{\#}}$1    &90.9861   &19.4012 \\
Unet${\rm{\#}}$2    &91.4635   &19.7707\\
Ynet                &87.2459   &17.8173\\
DEUnet              &90.6457   &19.3046\\
\hline
DuDoUnet                  &\bf{92.6300}    &\bf{20.4012}\\
DuDoUnet${\rm{ + }}$MI    &\bf{93.5622}    &\bf{20.8859}\\
\hline
\end{tabular}}
\end{table}

\section{Conclusion}
\label{sec:foot}

In this paper, we propose DuDoUnet for limited-view PA imaging reconstruction.
The specially designed ISB in DuDoUnet can greatly share dual domain information and distinguish artifacts.
Additional mutual information constraint makes encoder learn more prior knowledge to compensate limited-view information loss.
The experimental results showed that our method performed much better with higher SSIM and PSNR compared with other conventional methods.

%\section{Compliance with Ethical Standards}
%This is a numerical simulation study for which no ethical approval was required.

\bibliographystyle{IEEEbib}
\bibliography{strings,refs}

\end{document}